# Realization of advanced passive silicon photonic devices with subwavelength-grating structures developed by efficient inverse design

**Jingshu Guo, a,b,† Laiwen Yu, a,† Hengtai Xiang, a Yuqi Zhao, a Chaoyue Liu, a Daoxin Dai** a,b,c,*

aState Key Laboratory for Modern Optical Instrumentation, College of Optical Science and Engineering, International Research Center for Advanced Photonics, Zhejiang University, Zijingang Campus, Hangzhou, 310058, China

bJiaxing Key Laboratory of Photonic Sensing & Intelligent Imaging, Intelligent Optics & Photonics Research Center, Jiaxing Research Institute Zhejiang University, Jiaxing 314000, China

cNingbo Research Institute, Zhejiang University, Ningbo 315100, China

**Abstract**. The realization of ultra-compact passive silicon photonic devices is becoming more and more important for the future large-scale photonic integration as desired for many systems. Although some compact silicon photonic devices have been demonstrated by using inverse design, the device performance is still insufficient for real applications. Here, we propose and realize several representative ultra-compact advanced passive silicon photonic devices with decent performances by introducing subwavelength-grating (SWG) structures developed by our high-efficiency inverse design method. These devices are designed by optimally manipulating the multimode excitation and the multimode interference in a region defined with SWG structures. These SWG structures with excellent feature-size uniformity are more fabrication-friendly than those random nano-structures used in previous inverse-designed photonic devices. The high-efficiency of our inverse design method is attributed to a novel search-space-dimension control strategy and the efficient problem-oriented electromagnetic-field solvers available for SWG structures. Specifically, we present the realization of a 6-channel mode (de)multiplexer, a broadband 90°-hybrid, and a two-channel flat-top wavelength demultiplexer as some examples, which can hardly be realized by previously reported inverse design approaches. These devices exhibit ultra-compact footprints as well as decent performances when compared to the counterparts developed by the classical theory.

* E-mail: dxdai@zju.edu.cn
† Jingshu Guo and Laiwen Yu contributed equally to this work.

## 1 Introduction

Integrated photonics[1] has been known as a key technology of optoelectronics. By using on-chip light manipulations, the bulk optical and optoelectronic systems can be integrated to compact and robust photonic integrated circuits (PICs), bringing significant advantages in cost, size and power consumption. As the integration intensity of the PICs increases, researchers have been devoted to develop advanced passive silicon photonic devices with high performances and compact footprints.[2] For this problem, inverse design[3-6] provides a solution. An inverse design method translates the photonic design into finding the solution for a black-box problem, in which way the device structure is generated according to the given objective function by using computer aided design methods[7] (such as the evolutionary algorithms (EAs) or artificial neural networks (ANNs)[8]). Inverse design can break the limitation of the conventional design methods guided by classical theories, and thus potentially results in much more compact footprints in contrast to the conventional devices.[9,10]

For inverse design, the set of the to-be-decided geometry parameters is usually called as an "individual". An individual and its corresponding set of the objective-values consist of a "sample", which can be used for the training of ANNs. All the possible individuals constitute the high-dimension search space. Similarly, all the objective-value set constitute the solution space whose dimension is the objective-value set dimension. As it is well known, a multi-objective problem[11] is widely recognized as a big challenge in the field of computing mathematics. Currently, the inverse design of photonic devices is mainly limited to low-dimension solution spaces, in which case only a few objective-values are allowed to be involved. When designing a photonic device with a high-order scattering matrix, a large number of matrix elements should be optimized simultaneously, which makes the design optimization pretty difficult. For example, there are few results of demonstrating the inverse design of a 90° hybrid because the optimization is challenging when both amplitude and phase responses of all four output ports should be considered.[12] Similarly, the design complexity for mode (de)multiplexers increases significantly with the channel number and thus currently the mode (de)multiplexers developed with an inverse design still have quite limited mode-channels (no more than four).[12-14] Therefore, it is still very desired to develop an improved inverse design approach available for developing advanced passive silicon photonic devices with functional complexity.

To solve a black-box problem, the conventional technical route is utilizing optimization algorithms, such as the evolutionary algorithms (EAs),[15] the search algorithms (e.g., binary search algorithm[10]), and the gradient-based algorithms (e.g., gradient descent optimization[9], topology optimization[16]). The geometry framework is usually a complete black-box with a very high search space dimension (e.g., 100-1000[3,9,10,12-14]). The most popular geometry frameworks for the fully atomic inverse design are the complete black-box structures, such as no-shape-constraint topology optimization structures[6,9,14] and pixel structures,[10,12,13] both of which have to introduce a high search space dimension (in scale of $\sim 10^3$). In order to design application-oriented photonic integrated devices with decent performances, the multi-objective problems should be solved efficiently even when no reasonable initial individual is available. In this case, the optimization convergence usually becomes very difficult due to the high-dimensional search space and the high-dimensional solution space, and huge computational resources are needed. Moreover, the region to be designed is usually very limited (e.g., typically <10×10 μm²[6,9,10,12-14]) in order not to introduce an unacceptable search space dimension (e.g., >$10^4$). However, the device performance might not be sufficiently excellent for the real applications due to the limited solution space. As a result, the fully atomic inverse design usually cannot perform well for complex functional devices of photonics, e.g., the obtained performances may not meet the application requirements. From perspective of device fabrication, the device generated from the fully-automatic inverse design usually has a very tiny feature-size and the structure might be highly non-uniform, which makes the fabrication requirement very crucial. As an alternative, recently ANNs provide another technical route for inverse design.[8,17] It has been reported that a well-trained forward-modeling ANN may solve a nanophotonic particle inverse-design problem much more efficiently than conventional optimization algorithms.[18] However, the training of the ANN still requires a large amount of computing resources. In ref. [18], there are as many as 50,000 samples needed to train a forward-modeling ANN for an 8-sample-dimension problem. Therefore, the ANN inverse design mainly focuses on the specific grating design problems with low search space dimension of less than $10^{8,17}$ and consequently it is highly desired to reduce the dimension.[19,20]

As discussed in ref. [21], the issue for the fully-automatic inverse design may be solved with the prior knowledge and intuition of experienced designers, in which way the computational costs can be reduced by applying intelligent constraints and supplying an exceptional initialization. Previously, we proposed a universal manipulation strategy of on-chip multimode excitations/interferences by using all-dielectric metamaterial waveguides, and mode exchangers were successfully designed with low computation cost.[22] Fundamentally, various on-chip photonic devices for light manipulations with its amplitude,[23] wavelength,[24] polarization,[25] phase,[26] as well as mode[27] can be realized by appropriately manipulating the mode excitations and the mode interferences.

In this work, we propose a high efficiency inverse design approach for developing advanced passive silicon photonic devices based on branch waveguides with sub-wavelength grating structures. Supervised by the theory of multimode excitations/interferences,[22] this geometry framework enables high degree of freedom and low search space dimension of ~10-$10^2$ simultaneously. Importantly, it naturally supports a multi-stage optimization strategy with manual intervention. Benefiting from the experiences of traditional designs, the dimension of the search space increases minimally during the design process, which simultaneously enables fast convergence and high design flexibility. Meanwhile, some advanced optimization algorithm (like Covariance Matrix Adaptation Evolution Strategy: CMA-ES) and efficient electromagnetic (EM) simulation tools are applied to further improve the computation efficiency. Finally, three representative devices are designed and fabricated, confirming the effectiveness of the present approach for handling complex multi-objective problems. Furthermore, the designed photonic devices presented here are based on sub-wavelength grating structures and thus have good feature-size uniformity, which can be fabrication-friendly.

## 2  Results

*2.1 Theory and strategy*

For an inverse design, there are two key points to improve the design efficiency. First, the search space dimension should be minimized as much as possible. Second, the search space should be well defined so that the final optimal design is included. As a result, in this paper we propose an inverse design method with unique features, which are useful for efficiently developing advanced passive silicon photonic devices, as described below.

*2.1.1 Device geometry definition*

The device structure is basically defined geometrically with some waveguide branches separated by SWG metamaterial structures, as shown by the example given in Fig. 1(a). Definitely, the devices with more ports have more branches. Here each branch (see the deep-blue part in Fig. 1(a) consist of several segments longitudinally, and each segment is defined by its corner locations (i.e., the black pillars in Fig. 1(a)). As Fig. 1(a) shows, an SWG structure is usually defined by the period $P$ and the fill factor $\varLambda$, which can be preset or adjusted freely as desired. The SWG structures enable the dispersion engineering to realize broadband or wavelength-sensitive photonic devices when desired.[22,28] These to-be-designed geometry parameters are included in the individual for the optimization.[22]

This device geometry definition is derived from our previous work,[22] where a universal strategy of on-chip multimode excitation/interference was demonstrated. In this work, the optimization of the design region enables flexible mode manipulation and further realize on-chip light manipulations. The present design strategy is potentially useful as a universal approach for developing various optical functional devices even without any special initialization. This is different from those previously reported works based on multi-sectional MMIs optimized with appropriate initialization.[23,29] In particular, the multi-branch structure proposed here as the initialization can build up a relatively smooth connection between the input/output ports and the design region. As a result, it is possible to achieve excess losses (ELs) lower than those conventional inverse design devices with random nanostructures.

Apparently, the search space dimension is decided by the number of the lengthwise sections in the design region. Definitely any design region with $n$ sections can be redefined with $2n$ sections when needed. Such flexibility in the design makes it convenient to gradually adjust the search space dimension. For example, a low search space dimension with e.g., $n=10$ only can be adopted at the first stage of the optimization, so that one can achieve a sub-optimal solution with a low computation cost. The sub-optimal solution can be further used to initialize the second stage of optimization for which the search space dimension is improved to be with $n=\sim 10^2$. In this way, the present approach has much higher optimization efficiency than the conventional inverse design approaches.

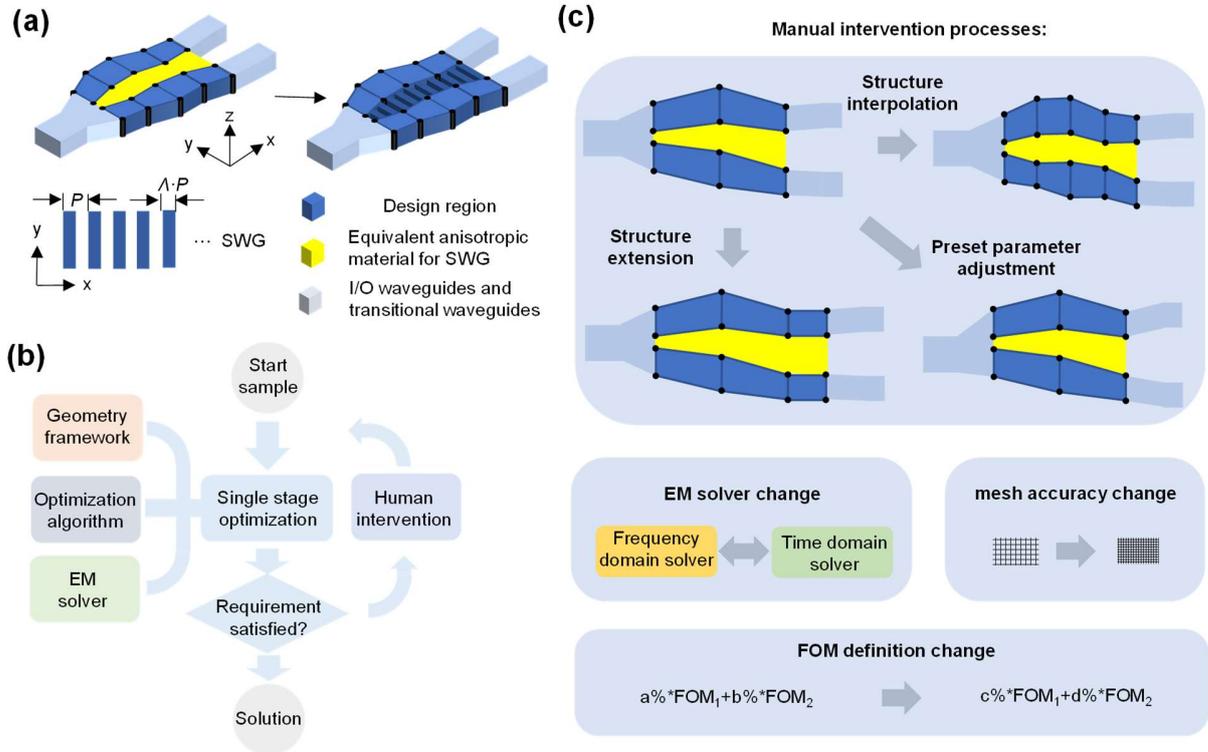

**Fig. 1** The proposed inverse design strategy for passive photonic devices. (a) The design framework demonstrated by an example of a photonic device with one input port and two output ports. Here the sub-wavelength grating (SWG) has a period of $P$ and a fill factor of $\Lambda$. b) The design flow chart with manual interventions. (c) The typical manual intervention operations.

*2.1.2. Iteration strategy with manual interventions*

Figure 1(b) shows the design flow chart of the present inverse design method, which is based on a multi-stage optimization iteration strategy with manual interventions. One stage of optimization is terminated when the FOM comes to a standstill. The corresponding solution is then used to initialize the next stage of optimization, and manual interventions may be introduced in the initialization, the EM solver setting, or other parameter setting. Figure 1(c) gives several kinds of typical manual intervention processes, which include the following cases. (*1*) The search space dimension can gradually be increased as the optimization stage iterates. For example, as Fig. 1(c) shows, a stage solution with 2 sections and 12 definition corners can be used to derive new generation with 4 sections and 20 definition corners. (*2*) When possible, one should first use the most efficient EM solver in the early stages of optimization and then use a high-precision EM solver (which is usually time-consuming) for the final stages of optimization. (*3*) For a multi-objective problem with a high-solution space dimension, the definition of the FOM may also be modified manually as the optimization stage iterates to balance the optimizations of different objective-values (Fig. 1(c)). (*4*) When no gratifying solution is obtained after several stages of optimizations, the global settings of the computation need to be modified. For example, the design region can be extended by the structure extension process. The preset parameters (like the input/output waveguide locations) can also be changed, as Fig. 1(c) shows.

Assisted with some manual interventions, the present inverse design with the multi-stage optimization strategy has high potentials to realize high-performance functional photonic devices since the search-space dimension for the final-stage optimization could be sufficiently high. Meanwhile, the optimization can be efficient because the geometry definition strategy is used to reduce the search-space dimension significantly.

*2.1.3. Optimization algorithm*

We use the Covariance matrix adaptation evolution strategy (CMA-ES),[30] which is known as an advanced optimization algorithm solving difficult black-box problems with large budgets. CMA-ES and its variants were found to outperform other optimization algorithms when the search space dimension is high (e.g., >20).[15,31] Currently, there have been few works using CMA-ES for the inverse designs of photonic integrated devices.[32] In this work, this advanced optimization algorithm is introduced to be helpful for efficient inverse designs. More details about this algorithm are given in **Materials and methods**.

*2.1.4. EM simulation tool*

To evaluate the FOM, the EM simulation should be carried for each individual. The typical EM simulation tools include Finite-Difference Time-Domain (FDTD),[10] Finite-Difference Frequency Domain (FDFD),[7] Finite Element Method (FEM),[33] and Eigenmode Expansion (EME).[34] The EM solving simulations usually occupy most of the computation resource in inverse designs. Apparently, the simulation time usually increases with the device footprint.

EME is also a frequency-domain algorithm whose computation speed and precision depend on the number of the eigenmodes involved and the number of the sections along the light propagation direction (rather than the device footprint).[35] With EME, the total scattering matrix for all the ports of the device can be achieved in one-run. In this case, only one run is needed even for a 6-channel mode DEMUX (with six output ports). In contrast, one has to run the FDTD simulation six times for the six mode-channels one by one. Furthermore, an EME simulation may be much faster than a 3D-FDTD simulation when the device has a large footprint (e.g., the length > 10 μm). However, the EME method does not work well when there are sharp bends or non-waveguide structures. Fortunately, the initialized structure is defined very well and thus intrinsically satisfies the requirement of EME.

Therefore, in this paper the EME method is introduced as an important tool for the simulation. In the early stages of optimizations, EME is used to obtain a preliminary optimization solution, which is then used as a good initialization for the following optimizations stages with a 3D-FDTD simulation. For example, when designing a 6-channel mode DEMUX, the EME simulation takes 80.6 hours only to improve the FOM from 10.94 to 0.42. The preliminary solution obtained by the EME simulation is already very close to the final optimization result. Definitely the introduction of efficient EME helps greatly to improve the efficiency of our inverse design method for developing high-performance optical functional devices.

In short, the present inverse design method combines the search-space-dimension-control strategy, the advanced optimization algorithm, and the flexible EM simulation strategy, so that it enables to solve efficiently multi-objective optimization problems. In the following sections, several advanced passive silicon photonic devices are presented by utilizing this inverse design method. In contrast, the design of these devices is very challenging due to the complexity when using conventional inverse design methods.

*2.2 Realization of a 6-channel mode (de)multiplexer*

A mode (de)multiplexer is the fundamental device in mode-division-multiplexing systems. Typically, high-performance mode (de)multiplexers are often realized by utilizing multi-stage asymmetric directional couplers.[36,37] Recently, ultra-compact mode (de)multiplexers have been demonstrated by using the inverse design methods.[12-14] However, these inverse-designed mode (de)multiplexers have no more than four channels. The reason is that the inverse design becomes much more difficult due to the increased complexity when it is desired to be with more channels (more ports).

Here we propose and demonstrate a 6-channel mode (de)multiplexer for the first time by utilizing the inverse design method. As Fig. 2(a) shows, the initial geometry framework is based on an ordinary 1×6 branch configuration without special considerations. In the optimization region, there are six polygon waveguides defined in the following way. Along the lengthwise direction, there are $n$ sections defined with the lengths $\boldsymbol{L} = [l_1, l_2, …, l_n]$. At the $yz$ cross-sectional interface between the ($n$-1)-th and the $n$-th sections, the parameters $\boldsymbol{D_n} = [d_{n1}, d_{n2}, …, d_{n12}]$ defines the lateral corner location in the $y$ direction (see Fig. 2(a)). The metamaterial SWG structure has a period $P$ and a fill factor $\Lambda$. Specifically, here we choose $P = 200$ nm and $\Lambda = 50\%$. For the input/output (I/O) waveguides, $w_{in}$ and $w_{out}$ respectively denote the waveguide widths, $d_{in0}$ and $d_{out0}$ respectively denote the positions of Port #1 and Port #7 in the $y$ axis, $d_{out}$ is the separation between the adjacent output waveguides. The parameter $l_{in}$ ($l_{out}$) denotes the length of the waveguide taper connecting the optimization region and the input (output) waveguide. The parameter set $\boldsymbol{S_{total}}$ to be optimized can combine flexibly any related parameters. For example, one might have $\boldsymbol{S_{total}} = [\boldsymbol{D_1}, \boldsymbol{D_2}…\boldsymbol{D_n}, \boldsymbol{D_{n+1}}]$ or $\boldsymbol{S_{total}} = [\boldsymbol{D_1}, \boldsymbol{D_2}…\boldsymbol{D_n}, \boldsymbol{D_{n+1}}, \boldsymbol{L}, l_{in}, l_{out}]$.

For the scattering matrix of a photonic device, the element $S_{ij}$ denotes the coupling coefficient from mode #$j$ to mode #$i$. For the 6-channel mode (de)multiplexer considered here, the bus waveguide at Port #1 supports six modes (i.e., the $TE_0$-$TE_5$ modes denoted as modes #1-#6). The output waveguides at Ports #2-#7 supports the corresponding $TE_0$ modes (denoted as modes #7-#12). Ideally, a 6-channel mode (de)multiplexer should have the transmissions of $|S_{ij}|^2=1$, where $|i-j|=6$, $1 \leq i \leq 6$ and $7 \leq j \leq 12$. The FOM for optimization is defined by $FOM = -\frac{1}{6}\sum_{i=1}^{6} 20\log_{10}|S_{(i+6)i}|$. Even though the optimization was run with a single operation wavelength of $\lambda = 1550$ nm, fortunately the optimized photonic device is very possible to work very well with a broad bandwidth, as demonstrated previously.[22]

Figure 2(b) shows the FOM and the total computation time cost as the iteration generation updates. The computation time was evaluated on an ordinary PC (details in **Materials and methods**). The milestone structures generated during the optimization are also depicted in Fig. 2(b). In terms of the fabrication process, the $SiO_2$-filling in the SWG slots usually depends on the $SiO_2$ deposition technology. In the design, we consider the cases with air slots or $SiO_2$ slots. The air-slot device was designed first with a regular linear initial structure (Fig. 2(b)). The EME method was used at the early stages, and the FOM reaches 0.42 dB after 384 generations of iteration (80.6 hours). The FDTD simulation with two types of mesh precision was then used in the following stages of optimization. Finally, the FOM is about 0.69 dB obtained with 104 generations (41.5 hours). This corresponding design was then used as the initialization for further optimizing the air-slot and SiO2-filled devices with high-precision FDTD simulation. As Fig. 2(b) shows, the optimization of the air-slot device finishes at the 507$^{th}$ generation with an FOM of 0.55 dB, costing 247.1 hours in total. And the optimization of the SiO2-filled device finishes at the 590$^{th}$ generation with an FOM of 0.62 dB, costing 419.2 hours in total (including the 122.1 hours for the optimization of the air-slot device initially). For such multi-stage designs, manual intervention processes were carried. For example, the section number $n$ changes from 6 to 24. More details are given in Section S1 of Supporting Information. The two types of devices have similar theoretical performances, providing excellent options for the labs/fabs with different fabrication technologies. In the following section, we focus on $SiO_2$-filled devices regarding to the fabrication technologies available.

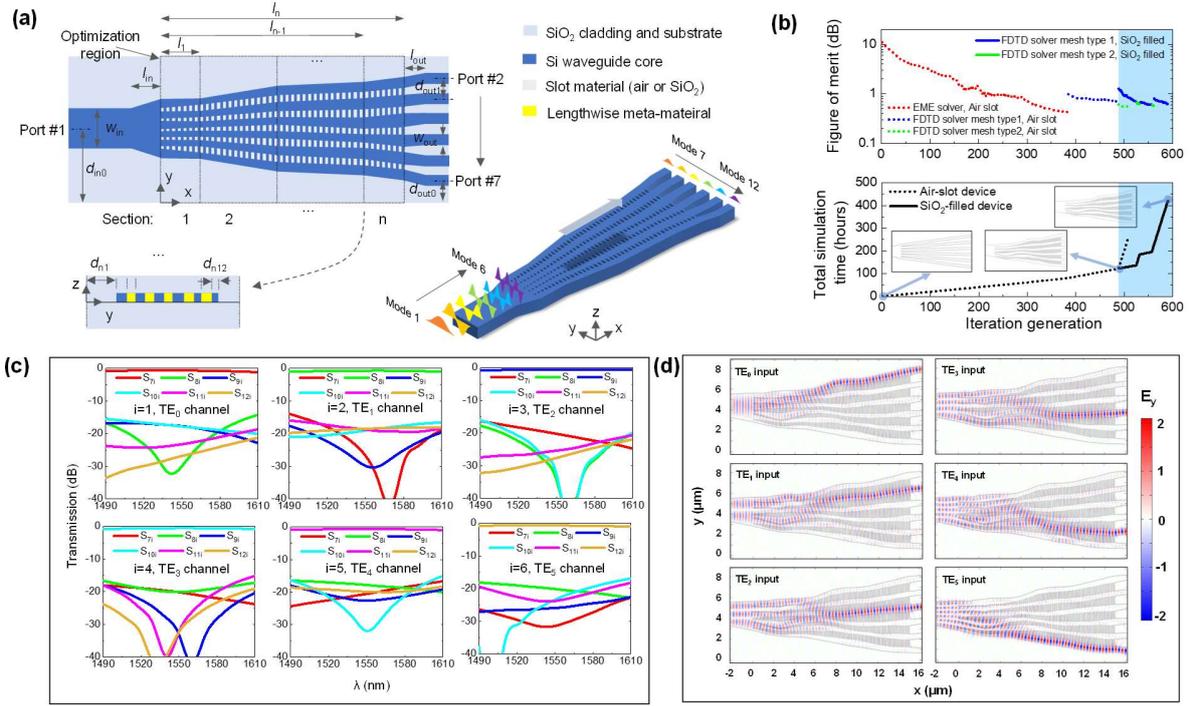

**Fig. 2** Inverse design of a 6-channel mode (de)multiplexer on silicon. (a) Top view (xy-plane), 3D view, and cross-sectionals view (yz-plane) at the interface between Section $n$-1 and Section $n$ (not in scale). (b) The figure of merit (FOM) and the simulation time cost as functions of the iteration generation for air-slot and $SiO_2$-slot devices. Inset: the milestone structures in the optimization flow. (c-d) The 3D-FDTD simulation performances of the designed $SiO_2$-filled mode (de)multiplexer including the transmissions (c) and the light propagation fields for different input modes at 1550 nm (d). Modes #1-#6 are respectively the $TE_0$- $TE_5$ modes supported by the bus waveguides (Port #1), and modes #7-#12 are respectively the $TE_0$ mode of Ports #2-#7. Here the silicon photonic waveguides with a 220-nm-thick silicon core and a $SiO_2$ upper-cladding are used.

According to the high-precision 3D-FDTD simulations with very fine meshes (26 mesh points per effective-wavelength scale), the designed $SiO_2$-filled device has a footprint as compact as 7.5×18 μm$^2$, low excess losses (ELs) of 0.53-1.03 dB, and low crosstalks (CTs) <−15.1 dB in the wavelength range of 1500-1600 nm for all mode channels, as Fig. 2(c) shows. In Fig. 2(d), the simulated light propagation at 1550 nm is given, showing impressive mode (de)multiplexing behaviors. The devices were fabricated by the regular processes from Applied Nanotools Inc.[38] (details in **Materials and methods**). Figure 3 shows the experimental results. Here the measured photonic integrated circuit (PIC) consists of a pair of mode (de)multiplexers and fiber-to-chip grating couplers (see Fig. 3(a)). Figure 3(b) shows the SEM image of the fabricated device and the normalized transmissions of the PIC are shown in Fig. 3(c). It can be seen that all channels for the fabricated mode (de)multiplexers have a low EL of <~1dB and low CT of <−10 dB in the wavelength range of 1520-1610 nm experimentally. The slight ripples in the transmission spectrums may be attributed to the non-zero reflection of the mode (de)multiplexers.

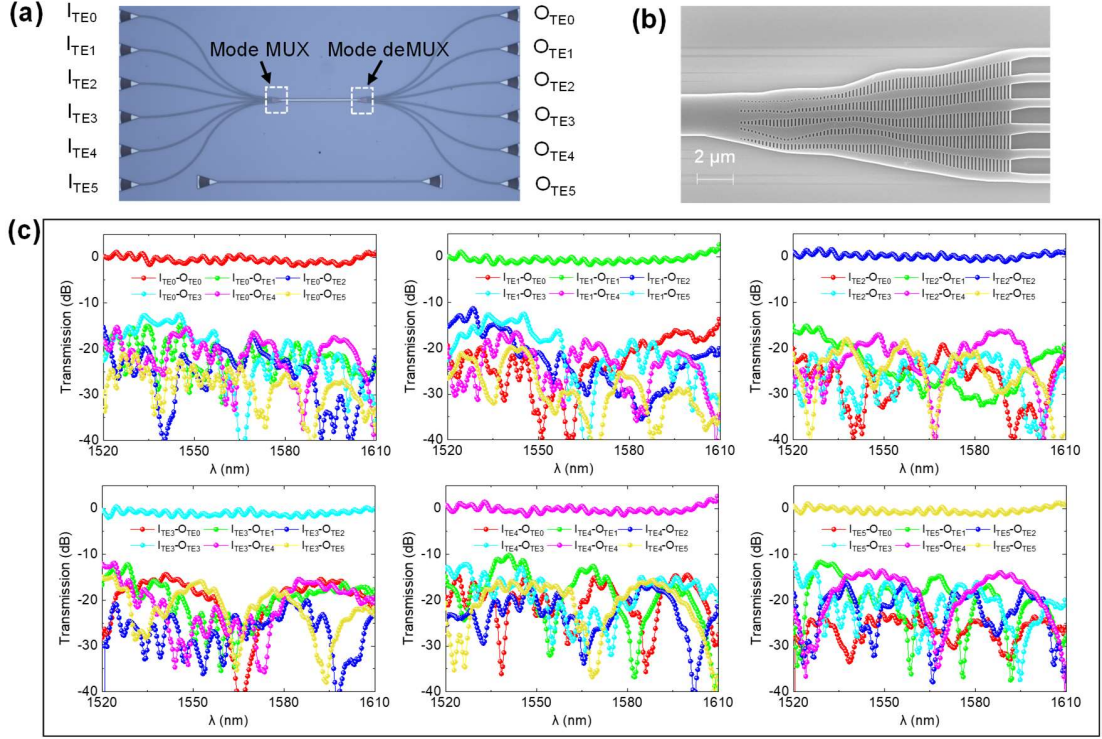

**Fig. 3** Fabricated devices and measured results. (a) Microscope picture for the fabricated silicon PIC consisting of a pair of mode (de)multiplexer with six input ports ($I_{TE0}$-$I_{TE5}$) and six output ports ($O_{TE0}$-$O_{TE5}$). (b) SEM image of a mode (de)multiplexer. c) Normalized transmissions of different port pairs.

*2.3 Realization of a 90° hybrid*

A 90° hybrid is the key device in optical coherent communication systems. Basically speaking, a 90° hybrid has two input ports (i.e., port #1 for the signal and port #2 for the local oscillator) and four output ports (#3, #4, #5 and #6), as shown in Fig. 4(a). Generally speaking, the optimal design of a 90° hybrid is pretty challenging because not only the amplitude responses but also the phase responses at the four ports should be considered.

For a 90° hybrid, the scattering matrix of the electric field amplitude should satisfy 8 requirements of $|S_{ij}|^2=1/4$ ($j$=1, 2; $i$=3, 4, 5, 6) while the phase differences between the output ports should satisfy the condition of $\{\theta_3-\theta_6, \theta_4-\theta_6, \theta_5-\theta_6\} = \{-90°, 90°, 180°\}$, where $\theta_i$=arg($S_{i2}$)-arg($S_{i1}$) ($i$=3, 4, 5, 6). In this problem, there are 11 target values, including 8 amplitude requirements and 3 phase requirements. One should notice that the correlation of these target values is weak in the optimal design of 90° hybrid. Specifically, the amplitude-target-value-set $\{|S_{ij}|^2, j=1\}$ for port #1, the amplitude-target-value-set $\{|S_{ij}|^2, j=2\}$ for port #2 and the phase-target-value-set [$\theta_3-\theta_6$, $\theta_4-\theta_6$, $\theta_5-\theta_6$] have almost no correlation among them. In this case, the optimization is often likely to be trapped locally and it is hard to reach the globally optimal design. Nevertheless, the present high-efficiency inverse design works well and gives a high-performance 90° hybrid successfully. As

shown by the details in Section S2 of Supporting Information, the FOM definition can be adjusted manually in the multi-stage optimization process to ensure globally optimization for all the target values.

Finally, the 90° hybrid is designed with a footprint of ~13×5 μm², and one has $\{\theta_3, \theta_4, \theta_5, \theta_6\} \approx \{0°, -180°, 90°, -90°+133°\}$. As a result, ports #3 and #4 are the in-phase (I) channels, while ports #5 and #6 are the quadrature (Q) channels. Figure 4(b) shows the simulated light propagation when the $TE_0$ modes with different phase differences $\triangle\theta_{21}$ are respectively injected into the two input ports. One can clearly see that the interference cancellation happens at one of the four output ports according to the phase difference. It proves that the present design method is effective even when both the amplitude and phase responses are required to be considered. Figure 4(c)-4(d) shows the simulated amplitude and phase responses in the wavelength range of 1530-1570 nm. It can be seen that the transmission losses are 6.6-7.4 dB and the ELs[29] of the signal- and LO-ports are less than 0.96 dB, while all the phase errors are within −4.6°~ 4.6°and the common mode rejection ratio (CMMR[29]) are >26.3 dB for all I/Q channels.

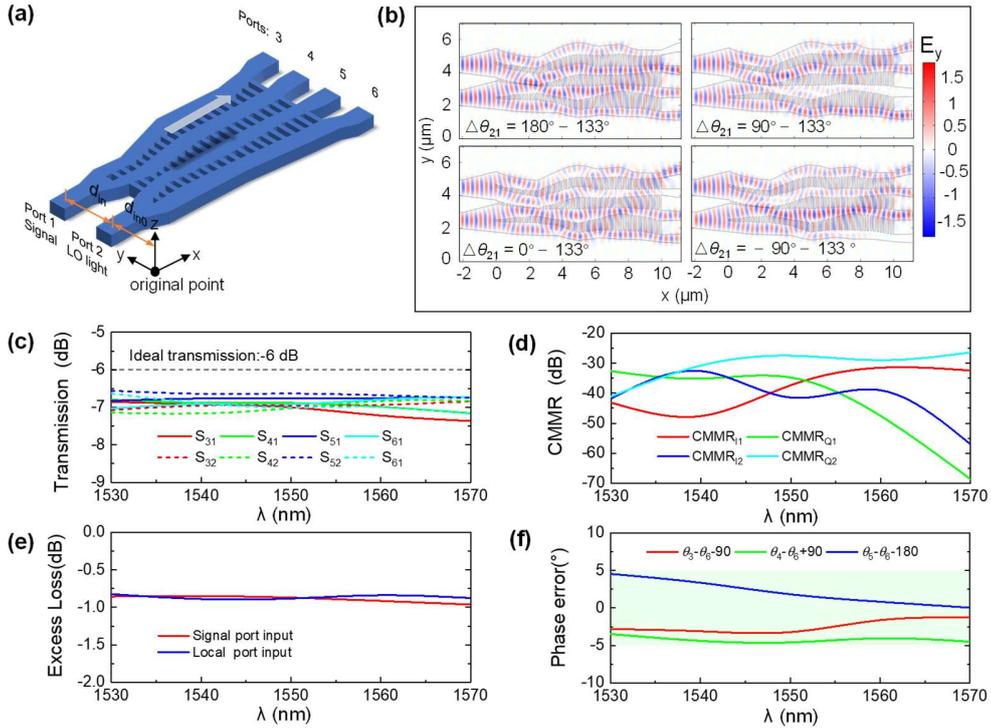

**Fig. 4** Inverse design of a 90° hybrid on silicon. (a) The 3D schematic diagram. LO: local oscillator. (b-f) The simulated performances of the final design (SiO$_2$-slot device): the simulated light propagation fields with varied phase difference $\triangle\theta_{21}$ between Ports 2 and 1 at 1550 nm (b), the transmissions (c) and CMMRs (d), the ELs (e) and phase errors (f). Here the transmission is given by the *S* parameter (i.e., S$_{ij}$ channel is given by 20log$_{10}$|S$_{ij}$|). CMMR: common mode rejection ratio.

Figure 5 shows the measurement results of the fabricated 90° hybrid. In our experiments, the transmissions were measured with the testing PIC (Fig. 5(a)), while the CMMRs and the ELs are extracted from the measured transmissions. As shown in Fig. 5(b), the measured transmissions are −4.8~−9.0 dB in the wavelength range of 1530-1570 nm. The CMMRs of both I/Q channels are below −10.2 dB, and the measured ELs are below 1dB, as shown in Fig. 5(c). The phase responses were measured by the testing PIC consisting by a 1×2 power splitter, a pair of asymmetric interference arms, a 90° hybrid, as well as grating couplers (see Fig. S1 in Section S2 of Supporting Information), which has been widely used before.[29] The measured transmissions at the four outports are presented in Fig. 5(d), and the extracted phase errors are given in Fig. 5(c), which shows that the phase error in the wavelength range of 1530-1570 nm is within the range of −14.90°-10.63°.

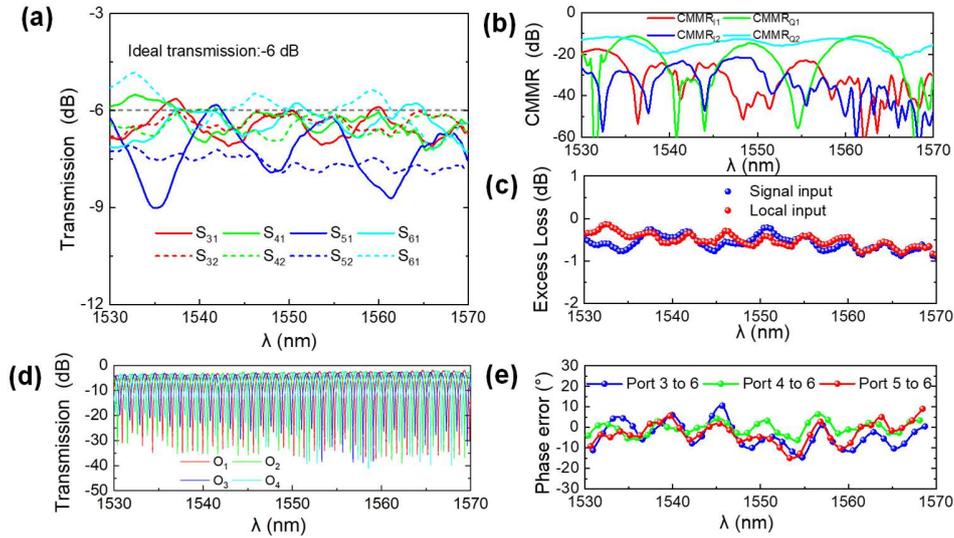

**Fig. 5** Experimental results of the fabricated 90° hybrid. (a) The measured transmissions. (b) The measured CMMRs. (c) The measured ELs. (d) The port-to-port optical transmission spectrums measured by the phase-test PIC. (e) The phase error extracted from the measured results of the phase-test PIC.

*2.4 Realization of a two-channel flat-top wavelength-division (de)multiplexer*

A wavelength (de)multiplexer[39] is a key element to separate/combine different wavelength channels in WDM systems. Here a two-channel wavelength (de)multiplexer is designed optimally with the present inverse design method. Currently, there have been several inverse designs for the wavelength (de)multiplexers with two channels,[9] three channels,[33,40,41] and six channels,[42] and most of them are narrowband. As it is well known, a flat-top response is often desired in practice because it becomes tolerant to some random wavelength variation. For the design of a wavelength (de)multiplexer with a flat-top spectral responses, the inverse design is usually challenging because it is required to include multiple objective values. In this example, the two wavelength-channels of 1290-1330 nm and 1470-1570 nm are considered.

Basically, the two-channel wavelength (de)multiplexer has an input port (#1) and two output ports (#2 and #3), as shown in Fig. 6(a). The 1550 nm channel and the 1310 nm channel are launched from port #1 and are then separated to port #2 and port #3, respectively. Here we choose several key wavelengths with a spacing of 20 nm for each channel, and the condition for the optimization is given by $|S_{21}(\lambda_a)|^2=1$, $\lambda_a \in \{1470, 1490, 1510, 1530, 1550, 1570\}$ nm and $|S_{31}(\lambda_b)|^2=1$, $\lambda_b \in \{1290, 1310, 1330\}$ nm. The optimization costs 302.5 hours (see more details in Section S3 of Supporting Information). Finally, the designed device has a compact footprint of $3.1 \times 12.5$ μm$^2$. Figure 6(b) shows the simulated light propagation for the two central wavelengths by using the 3D-FDTD method. As shown in Fig. 6(c), the designed wavelength (de)multiplexer has ELs<1 dB and ERs > 10 dB in the wavelength ranges of 1250-1330 nm and 1470-1610 nm. The ELs at 1310 nm and 1550 nm are respectively 0.67 dB and 0.36 dB, while the corresponding ERs are respectively 15.02 dB and 30.33 dB. The fabricated wavelength (de)multiplexer (see the inset in Fig. 6(d)) were then measured with the help of the PICs with different grating couplers working at the corresponding wavelength-bands of 1310 nm and 1550 nm. From the measured transmissions shown in Fig. 6(d), it can be seen that the EL is <1dB and the ER is >10 dB in the wavelength ranges of 1256-1344 nm and 1482-1591 nm.

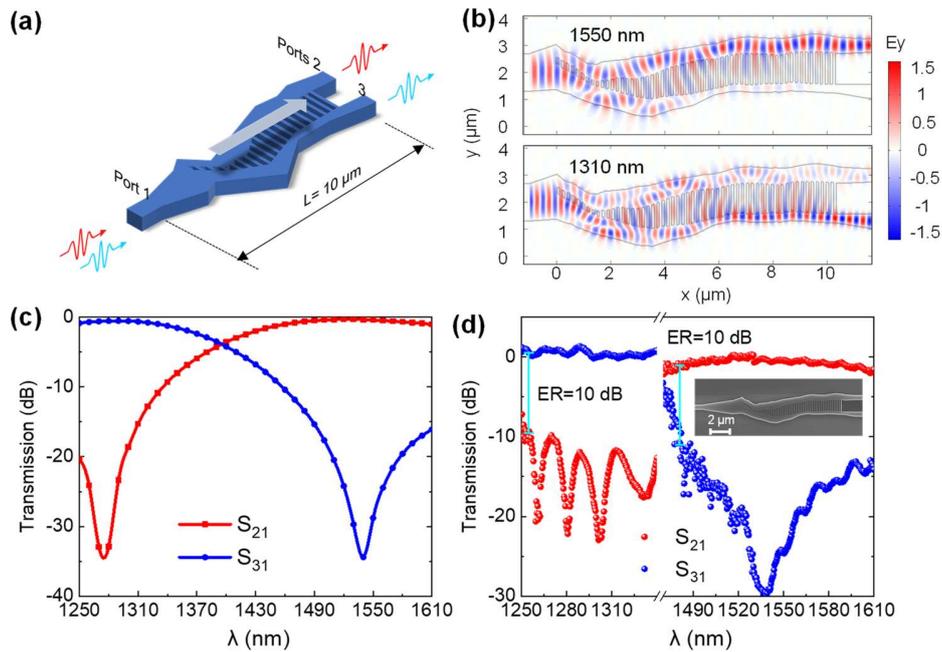

**Fig. 6** The inverse-designed two-channel wavelength multiplexer on silicon. (a) The 3D schematic diagram. (b) The simulated light propagation in the designed devices. (c) The calculated light transmissions. (d) The measured transmissions. inset: The SEM image of the fabricated wavelength demultiplexer.

*2.5 Comparisons and perspective*

Table 1 given a summary of various related silicon photonic devices reported. The state-of-the-art devices developed by classical theory design methods and inverse design methods are listed. Here, the devices are evaluated by the performances within the whole bandwidth except those non-flat-top wavelength (de)multiplexers and those labelled with "@ $\lambda_0$". Here "@ $\lambda_0$" is for the performance at the central wavelength $\lambda_0$. The classical theory design methods usually have advantages in device performances and the design scalability. For these devices, however, the difficulty for footprint shrinking is the key issue. In contrast, inverse designs have natural advantages in footprint compactness, while the bottleneck is the design efficiency for complex functional devices with a large number of to-be-optimized objective values, such as multi-channel devices or flat-top wavelength-selective devices. As Table 1 presents, the present mode (de)multiplexer exhibits high theoretical performances of low ELs, low CT, and large bandwidths comparable to or even better than those conventional mode (de)multiplexers.[37,43] The measured CT for the fabricated mode (de)multiplexers is slightly inferior to conventional mode (de)multiplexers, and the footprints depend on the channel numbers. More recently, a four-channel mode (de)multiplexers using tilt waveguide junctions was reported with footprint of 2×50 $\mu m^2$,[43] which represent the state-of-the-art level of the conventional methods. Definitely the device footprint will increase greatly if the channel number increases from 4 to 6. Furthermore, the performance degradation may occur as usual. To the best of our knowledge, this paper presents the first 6-channel mode (de)multiplexers developed with inverse design, owing to the high efficiency of our inverse design strategy. Moreover, our devices have advantages in the performances of the EL and the bandwidth when compared to the other inverse design counterparts.

For conventional 90° hybrids, the state-of-the-art is with an EL below 0.5 dB, a CMMR over 30 dB, and phase errors below 3° in the C band (1530-1565 nm) while the footprint is 21.6×27.9 $\mu m^2$.[29] In contrast, the present 90° hybrid has a footprint shrunk by ~10 folds and similar performances, including low ELs, high CMMRs, low phase errors, and large bandwidth. Our present 90° hybrid has shown much better performances on the EL, the imbalance, the phase error, and the bandwidth than the inverse-designed 90°-hybrid with two-dimensional code structures.[12] The broadband characteristic of our 90° hybrid also reflects the high-efficiency capability of our approach for the optimization with a large number of objective values (e.g., with multiple wavelengths involved).

Nowadays, there have been many wavelength (de)multiplexers designed conventionally,[39] for which high performances and footprint compactness are usually conflicted. As shown in Table 1, the two-channel flat-top wavelength (de)multiplexers designed with conventional methods are usually based on Mach-Zehnder interferometers,[44] waveguide Bragg gratings,[45] or wavelength-selective waveguide coupler,[46] and they have lengths in scale of ~$10^2$ $\mu m$. In contrast, the inverse-designed wavelength (de)multiplexer in this work has ultra-compact footprint of 3.07×12.46 μm$^2$. In recent years, there have been several inverse-designed wavelength (de)multiplexers.[9,33,40,41,42] However, they most are non-flat-top.[33,41,42] The inversed-designed wavelength (de)multiplexer in ref. 9 is indeed flat-top while the EL is higher than 1.8/2.4 dB for the central wavelengths of 1.31/1.55 μm. In contrast, our device has flat-top responses as well as a low EL of <1 dB. More recently, a three-channel flat-top wavelength (de)multiplexer was reported with a bandwidth of ~30 nm for an EL of ~1dB and an ER of >15 dB[40]. In contrast, our wavelength (de)multiplexer in this paper enables flat-top responses with an ultrabroad bandwidth of 80-110 nm. Furthermore, the present devices can be scaled easily for more channels by structure cascading when needed.

As Table 1 shows, the experimental results for the CT/ER/CMMR/IM are usually inferior to the simulation ones mainly due to the fabrication errors. In our case, all the geometry parameters including $D_n$=[$d_{n1}$, $d_{n2}$, …, $d_{n12}$] and the SWG feature size are chosen to be sufficiently large (e.g., >80 nm) according to the fabrication technology. Unfortunately, there exist some acute-angle structures at some corners, which leads to some performance degradation for the fabricated devices. In the future work, the device performance could be possibly improved by introducing smoothened subwavelength structures. The optimization efficiency should be improved further for realizing the devices with very high complexity such as the mode/wavelength (de)multiplexers with many channels.

**Table 1** Summary of the state-of-the-art mode (de)multiplexers, 90° hybrids, and wavelength (de)multiplexers on silicon.

| Device | Type, year | Footprint (μm$^2$) | EL (dB) | | Crosstalk/Extinction ratio/CMMR/IM (dB) | | Bandwidth(nm)[b] | | Special indicators [c] | | Ref. |
|---|---|---|---|---|---|---|---|---|---|---|---|
| | | | Sim. | Exp. | Sim. | Exp. | Sim. | Exp. | Sim. | Exp. | |
| Mode (de)multiplexer | Dual-core adiabatic tapers, 2018 | ~33×471 | NA | <1.8 | NA | CT<-14 | 140 | 80 | 10 Channels (5TE+5TM) | | [37] |
| | Tilt waveguide junctions, 2022 | 2×50 | <1 | <1.29 | CT<-17.4 | CT<-14.4 | 60 | 60 | 4 Channels | | [43] |
| | ANN-Inverse Design, 2021 | 2×17.5 | <1.1@1.55 μm[a] | <2.4@1.55 μm | CT<-10 | CT<-10 | 90 | 70 | 3 Channels | | [47] |
| | Inverse Design, 2020 | 5.4×6 | <2 (~1@1.55 μm) | <1.5 | CT<-18 | CT<-16 | 60 | 60 | 4 Channels | | [13] |
| | Inverse Design, 2021 | 6.5×6.5 | NA | 0~10 | NA | CT<-14 | NA | 60 | 4 Channels | | [14] |
| | Inverse Design, 2022 | 4.8 × 4.8 | <5 (<1.1@1.55 μm) | <2.5 | CT<-12 | CT<-12 | 100 | 40 | 4 Channels | | [12] |
| | Inverse Design | 7.5×18 | <1.03 | <~1 | CT<-15.12 | CT<-10 | 100 | 90 | 6 Channels | | This work |
| 90° hybrid | MMIs + Phase shifter, 2017 | 21.6×27.9 | <0.45 | <0.5 | CMMR>30 | CMMR>30 | 35 | 35 | - | PE<3 | [29] |
| | Inverse design, 2022 | 4.8×4.2 | <2 (<0.5@1.55 μm) | <2.1[a] | -2~2.4(IM) | ~±2(IM) | 100 | 40 | PE<70 (<6.5@1.55 μm) | PE<27 (<10@1.55 μm) | [12] |
| | Inverse Design | 4.71×13.03 | <0.96 | <1 | CMMR>26.33 IM<0.37 | CMMR>10.2 IM=-2.82~1.5 | 40 | 40 | PE<4.6 | PE<14.9 (<5@1.55 μm) | This work |
| WDM | Inverse design, 2015 | 2.8×2.8 | ~1.7@1.3/1.55 μm | >1.8/2.4 | ER>~13 | ER>11 | 100/170 | 100/170 | 2 Channels, Flat-top | | [9] |

| | | | | | | | | | |
|---|---|---|---|---|---|---|---|---|---|
| Inverse design, 2018 | 5.5×4.5 | 1.56/1.68/1.35 | 2.82/2.55/2.29 | ER>15 | ER>10.7 | NA | NA | 3 Channels, Non-flat-top | [41] |
| Inverse design, 2019 | 1.4×1.8 | 0.36/0.09/0.76 | 1.87/1.49/3.47 | ER>6.23 | ER>8.51 | NA | NA | 3 Channels, Non-flat-top | [33] |
| Inverse design, 2020 | 2.8×2.8 | 0.3/0.54 | NA | ER>15.29 | NA | NA | NA | 2 Channels, Non-flat-top | [42] |
| | 4.6×2.8 | ~1.9 | | ER=~13 | | | | 4 Channels, Non-flat-top | |
| | 6.95×2.8 | 0.31~2.12 | | ER>15.86 | | | | 6 Channels, Non-flat-top | |
| Inverse design, 2022 | 6.2 × 5.4 | ~0.8 | 1.2 | ER>17 | ER>15 | >30 | >30 | 3 Channels, Flat-top | [40] |
| Inverse Design | 3.07×12.46 | <1 | <1 | ER>10 | ER>10 | 80/140 | 88/109 | 2 Channels, Flat-top | This work |

Notes:

a) The superscript '@ $\lambda_0$' denotes that the result is only for the central wavelength $\lambda_0$ (similarly hereinafter).

b) For the wavelength (de)multiplexer, the bandwidth of each channel of the flat-top devices are given. For ref. 9, the simulated bandwidth condition is EL<4 dB and ER>~13 dB, while the experimental bandwidth condition is EL<5.4 dB and ER>11 dB. In our work, the bandwidth is for achieving EL<1 dB and ER>10 dB.

c) Here the special indicators are listed including simulated and experimental results, e.g., the channel numbers of mode (de)multiplexer and wavelength (de)multiplexer, the phase error of 90° hybrid, the flat-top feature.

d) Abbreviation definitions: EL, excess loss; CT, crosstalk; ER, extinction ratio; CMMR, common mode rejection ratio; IM, imbalance; PE, phase error; Sim., simulated result; Exp., experimental result; NA, not available; PBS, polarization beam splitter; WDM, wavelength (de)multiplexer.

## 3  Conclusion

In summary, we have proposed a high-efficiency inverse design method for developing advanced passive silicon photonic devices with high performances and compact footprints. Multi-stage optimizations with manual intervention processes are carried based on segmented geometry-definition assisted by SWG structures. Thanks to the manual intervention processes, the search space dimension increases gradually under control to realize fast convergence and high design degrees of freedom simultaneously. Meanwhile, the flexible setting of EM simulation tools and the efficient optimization algorithm CMA-ES further enable a high design efficiency for our method. Using this method, three types of advanced passive silicon photonic devices have been demonstrated, including a 6-channel mode (de)multiplexer, a 90° hybrid, and a two-channel flat-top wavelength (de)multiplexer. In contrast to the conventional inverse design methods, our method has the advantage in handling multi-objective problems and shows great potential to be applied widely for various advanced passive photonic devices. For instance, it enables the realization of the first 6-channel mode (de)multiplexer designed by inverse design method to the best of our knowledge. The first *broadband* 90° hybrid has also been realized with our inverse design, which is very challenging because there are 55 objective values involved. Furthermore, the present devices usually have low loss of <1dB, enabling the possibility of device cascading when needed. The demonstrated devices show decent theoretical and experimental performances comparable to their state-of-the-art counterparts designed conventionally. Meanwhile, their footprints are reduced greatly by 2-10 folds. The present work can be extended easily to III-V and lithium niobate systems. As a summary, this inverse design method is very helpful for the development of advanced passive photonic devices with high performance, design universality and footprint compactness, which may be useful for realizing next generation high-intensity photonic integrated circuits. In the future, further efforts in design strategy and computation algorithm should be made to achieve performance improvements and footprint compressions.

**Materials and methods**

*Covariance matrix adaptation evolution strategy (CMA-ES)*

CMA-ES is based on a multi-individual search in which the search individuals of each generation are sampled from multi-variate normal distributions. The individuals with good FOMs are used for guiding the evolution of the normal distribution parameters. The introduction of normal distribution follows the maximum entropy principle, making this algorithm powerful. The algorithm flow is described as follows. The individuals of each generation are obtained by sampling of the normal distribution $N(m, (\sigma^{(g)})^{2(c)}) \sim m^{(g)} + \sigma^{(g)} N(0, C^{(g)})$, where $m^{(g)}$ is the mean value of the search distribution at generation g, $\sigma^{(g)}$ is the step-size, and $C^{(g)} \in \mathbf{R}_{n \times n}$, is the covariance matrix of the normal distribution $N(0, C^{(g)})$ at generation g. The FOM of each individual is evaluated by EM solving and then ranked. After then, part of the individuals with good FOMs (usually the better 50%) are used for updating the parameters [$m^{(g+1)}$, $\sigma^{(g+1)}$, $C^{(g+1)}$] by covariance matrix adaptation algorithm for next-generation computation (details in ref. 30). As the generation increases, the population (including the mean value and the sampled individuals) will approach to the promising area covering good solutions. CMA-ES has been known as one of the most efficient evolutionary approaches, which is effective for the non-separable, ill-conditioned black-box problems with high dimensionality. In this work, the start sample is assigned to $m^{(1)}$, the initial step size $\sigma^{(1)}$ is set to 2~7 nm, depending on the specific problems, $C^{(1)}$ is an identity matrix.

*Electromagnetic field numerical simulation details*

The 3D-FDTD simulations and EME simulations were performed by using Lumerical FDTD Solutions and Mode Solutions. In this work, the simulation time was test on a personal computer (CPU: Inter Core i7-11700 @2.5 GHz, RAM: 64 G). The EME solver was carried with configuration of 1 process and 16 threads. The SWG metamaterial is equivalent to homogeneous anisotropic metamaterials.[22] The FDTD solver was carried with configuration of 4 processes and 1thread. The final performance confirmations of the devices were performed with dense meshes (i.e., mesh accuracy index of 6 or 8, corresponding to 26 or 34 mesh points per effective-wavelength scale).

*Device fabrication*

All the devices were fabricated on a silicon-on-insulator (SOI) wafer with a 220 nm-thick top-silicon layer and a 2 μm-thick buried dioxide layer. The silicon photonic waveguides were patterned by the processes of electron-beam lithography and inductively-coupled plasma dry-etching. Then the 2.2-μm-thick silicon dioxide cladding was deposited by using chemical vapor deposition.

**Disclosures**

The authors declare no conflicts of interest.

**Acknowledgments**

This work was supported by National Major Research and Development Program (No. 2018YFB2200200); National Science Fund for Distinguished Young Scholars (61725503); National Natural Science Foundation of China (NSFC) (62175216, 61961146003, 91950205); Zhejiang Provincial Natural Science Foundation (LR22F050001); The Fundamental Research Funds for the Central Universities; The Leading Innovative and Entrepreneur Team Introduction Program of Zhejiang (2021R01001).

**Code, Data, and Materials Availability**

Supporting Information is available. The data that support the findings of this study are available from the corresponding author upon reasonable request.

# Supporting Information

# Realization of advanced passive silicon photonic devices with subwavelength-grating structures developed by efficient inverse design

**Jingshu Guo, [a,b,†] Laiwen Yu, [a,†] Hengtai Xiang, [a] Yuqi Zhao, [a] Chaoyue Liu, [a] Daoxin Dai [a,b,c,\*]**

[a]State Key Laboratory for Modern Optical Instrumentation, College of Optical Science and Engineering, International Research Center for Advanced Photonics, Zhejiang University, Zijingang Campus, Hangzhou, 310058, China
[b]Jiaxing Key Laboratory of Photonic Sensing & Intelligent Imaging, Intelligent Optics & Photonics Research Center, Jiaxing Research Institute Zhejiang University, Jiaxing 314000, China
[c]Ningbo Research Institute, Zhejiang University, Ningbo 315100, China
[\*]E-mail: dxdai@zju.edu.cn
[†] Jingshu Guo and Laiwen Yu contributed equally to this work.

**Section S1. The details of the inverse-designed 6-channel mode (de)multiplexer**

In Table S1, the design flow details are given. For each stage, we present the information of the electromagnetic (EM) solver, optimization data-set type, section number, as well the sample dimension which is the dimension of the optimization data-set setting. In the whole design flow, the optimization region length is fixed to 15 μm, and the *FOM* definition stays the same. The *FOMs* of the initial individual and the optimized result of each stage are given in Table S1. The computation cost details are also shown in Table S1. The total computation time of each stage is the product of generation population, simulation time of one- individual, and optimization generation.

The air-slot device was successfully designed with 6 optimization stages and total computational time of 247.1 hours. The Stage-4 optimization result of the air-slot device was used as initial individual to design the $SiO_2$-filled device. The $SiO_2$-filled device was successfully designed within extra 5 stages of optimization costing 297.1 hours. In Table S2, the detail geometric parameters of the final designed $SiO_2$-filled 6-channel mode (de)multiplexer is given.

## Table S1 The details of the 6-channel mode (de)multiplexer design flow.

| Device Type | Stage | EM Solver (mesh type)[a] | Optimization data-set type[b] | Section number | Sample size | FOM (dB) Start | FOM (dB) End | Generation population | One-individual time (hour)[c] | Generations | Stage time (hour) |
|---|---|---|---|---|---|---|---|---|---|---|---|
| Air-slot | 1 | EME | 1[†] | 6 | 84 | 10.936 | 0.822 | 17 | 0.012 | 291 | 59.4 |
| | 2 | EME | | 12 | 156 | 0.822 | 0.424 | 19 | 0.012 | 93 | 21.2 |
| | 3 | FDTD(1)[*] | | | | 0.977 | 0.801 | | 0.021 | 33 | 13.2 |
| | 4 | FDTD(1) | 2 | 12 | 167 | 0.801 | 0.691 | | 0.021 | 71 | 28.3 |
| Air-slot | 5 | FDTD(2) | 2 | 12 | 167 | 0.610 | 0.577 | 19 | 0.323 | 6 | 36.8 |
| | 6 | FDTD(2) | | 24 | 323 | 0.577 | 0.548 | 21 | 0.323 | 13 | 88.2 |
| SiO$_2$-filled | 5 | FDTD(1) | 2 | 12 | 167 | 1.274 | 0.697 | 19 | 0.021 | 37 | 14.8 |
| | 6 | FDTD(2) | | 24 | 323 | 0.666 | 0.620 | 21 | 0.323 | 7 | 47.5 |
| | 7 | FDTD(1) | | | | 0.674 | 0.603 | | 0.021 | 25 | 11.0 |
| | 8 | FDTD(2) | | | | 0.589 | 0.578 | | 0.323 | 4 | 27.1 |
| | 9 | FDTD(2) | | | | 0.807 | 0.623 | | 0.323 | 29 | 196.7 |

Notes:
a) For the mesh of FDTD solver, Type 1: 10 mesh points per effective-wavelength scale, Type 2: 14 mesh points per effective-wavelength scale.
b) Type 1: the data-set is defined as $S_{total}=[D_1, D_2…D_n, D_{n+1}]$, the length array is given by $L=[l_1,l_2…l_n]= [1,2,…,n]·(L_n/n)$, $L_n$ = 15 μm. Type 2: the data-set is defined as $S_{total}=[D_1, D_2…, D_n, D_{n+1}, l_1, …,l_{n-1}]$, $L_n$ =15μm.
c) One-individual time was measured on the personal computer with performance parameters given in **Methods** of main text. For EME and FDTD methods, simulation of one sample needs one run and six runs of solvers, respectively.

## Table S2 The detail geometric parameters of the final designed SiO$_2$-filled 6-channel mode (de)multiplexer (unit: μm).

| $d_{mn}$ | | Optimization Region Parameters in lateral direction $n$ | | | | | | | | | | | | Optimization Region Parameters in lengthwise direction | | Else parameters | |
|---|---|---|---|---|---|---|---|---|---|---|---|---|---|---|---|---|---|
| | | 1 | 2 | 3 | 4 | 5 | 6 | 7 | 8 | 9 | 10 | 11 | 12 | | | | |
| | 1 | 2.844 | 0.195 | 0.093 | 0.435 | 0.158 | 0.17 | 0.16 | 0.493 | 0.163 | 0.436 | 0.089 | 0.408 | | | | |
| | 2 | 2.694 | 0.199 | 0.168 | 0.364 | 0.161 | 0.261 | 0.243 | 0.453 | 0.152 | 0.401 | 0.096 | 0.435 | $l_1$ | 0.655 | $w_{in}$ | 2.3 |
| | 3 | 2.545 | 0.159 | 0.278 | 0.266 | 0.164 | 0.406 | 0.36 | 0.395 | 0.142 | 0.369 | 0.083 | 0.486 | $l_2$ | 1.246 | $d_{in0}$ | 4.4 |
| | 4 | 2.418 | 0.15 | 0.303 | 0.208 | 0.165 | 0.5 | 0.412 | 0.373 | 0.158 | 0.449 | 0.078 | 0.562 | $l_3$ | 1.869 | $w_{out}$ | 0.5 |
| | 5 | 2.301 | 0.145 | 0.351 | 0.145 | 0.139 | 0.629 | 0.463 | 0.304 | 0.152 | 0.478 | 0.107 | 0.637 | $l_4$ | 2.543 | $d_{out0}$ | 0.9 |
| | 6 | 2.258 | 0.196 | 0.328 | 0.224 | 0.171 | 0.603 | 0.45 | 0.332 | 0.214 | 0.44 | 0.098 | 0.578 | $l_5$ | 3.142 | $d_{out}$ | 1.4 |
| | 7 | 2.229 | 0.294 | 0.275 | 0.353 | 0.181 | 0.53 | 0.387 | 0.321 | 0.249 | 0.424 | 0.106 | 0.55 | $l_6$ | 3.750 | $l_{in}$ | 2 |
| | 8 | 2.202 | 0.31 | 0.289 | 0.418 | 0.245 | 0.472 | 0.292 | 0.379 | 0.266 | 0.373 | 0.16 | 0.491 | $l_7$ | 4.382 | $l_{out}$ | 1 |
| | 9 | 2.167 | 0.372 | 0.27 | 0.482 | 0.357 | 0.406 | 0.2 | 0.413 | 0.298 | 0.376 | 0.197 | 0.414 | $l_8$ | 5.009 | | |
| | 10 | 2.06 | 0.384 | 0.344 | 0.514 | 0.35 | 0.378 | 0.261 | 0.361 | 0.373 | 0.308 | 0.288 | 0.402 | $l_9$ | 5.629 | | |
| | 11 | 1.946 | 0.422 | 0.407 | 0.577 | 0.382 | 0.337 | 0.281 | 0.333 | 0.45 | 0.285 | 0.401 | 0.398 | $l_{10}$ | 6.263 | | |
| | 12 | 1.823 | 0.454 | 0.496 | 0.538 | 0.345 | 0.352 | 0.315 | 0.344 | 0.499 | 0.339 | 0.511 | 0.436 | $l_{11}$ | 6.889 | | |
| $m$ | 13 | 1.723 | 0.499 | 0.58 | 0.496 | 0.298 | 0.322 | 0.354 | 0.35 | 0.595 | 0.42 | 0.632 | 0.444 | $l_{12}$ | 7.507 | | |
| | 14 | 1.589 | 0.578 | 0.566 | 0.456 | 0.332 | 0.303 | 0.415 | 0.391 | 0.645 | 0.446 | 0.642 | 0.487 | $l_{13}$ | 8.134 | | |
| | 15 | 1.475 | 0.585 | 0.563 | 0.398 | 0.385 | 0.322 | 0.48 | 0.384 | 0.747 | 0.492 | 0.652 | 0.489 | $l_{14}$ | 8.743 | | |
| | 16 | 1.347 | 0.644 | 0.538 | 0.367 | 0.454 | 0.353 | 0.567 | 0.446 | 0.705 | 0.453 | 0.662 | 0.45 | $l_{15}$ | 9.402 | | |
| | 17 | 1.268 | 0.682 | 0.527 | 0.317 | 0.486 | 0.381 | 0.626 | 0.49 | 0.694 | 0.452 | 0.648 | 0.427 | $l_{16}$ | 10.010 | | |
| | 18 | 1.178 | 0.621 | 0.595 | 0.328 | 0.535 | 0.415 | 0.727 | 0.445 | 0.623 | 0.464 | 0.683 | 0.441 | $l_{17}$ | 10.620 | | |
| | 19 | 1.068 | 0.574 | 0.669 | 0.359 | 0.574 | 0.414 | 0.793 | 0.471 | 0.587 | 0.492 | 0.716 | 0.435 | $l_{18}$ | 11.230 | | |
| | 20 | 0.959 | 0.534 | 0.699 | 0.381 | 0.697 | 0.395 | 0.841 | 0.504 | 0.586 | 0.499 | 0.739 | 0.478 | $l_{19}$ | 11.871 | | |
| | 21 | 0.853 | 0.544 | 0.743 | 0.369 | 0.776 | 0.382 | 0.887 | 0.535 | 0.603 | 0.51 | 0.751 | 0.511 | $l_{20}$ | 12.484 | | |
| | 22 | 0.792 | 0.539 | 0.75 | 0.45 | 0.777 | 0.43 | 0.844 | 0.556 | 0.668 | 0.46 | 0.795 | 0.542 | $l_{21}$ | 13.110 | | |
| | 23 | 0.728 | 0.535 | 0.751 | 0.528 | 0.788 | 0.464 | 0.818 | 0.589 | 0.76 | 0.426 | 0.837 | 0.541 | $l_{22}$ | 13.745 | | |
| | 24 | 0.713 | 0.528 | 0.707 | 0.555 | 0.789 | 0.581 | 0.764 | 0.609 | 0.796 | 0.464 | 0.842 | 0.603 | $l_{23}$ | 14.374 | | |
| | 25 | 0.716 | 0.519 | 0.702 | 0.564 | 0.806 | 0.655 | 0.712 | 0.65 | 0.82 | 0.466 | 0.818 | 0.654 | $l_{24}$ | 15.000 | | |

**Section S2. The details of the inverse-designed 90° hybrid**

In the design flow, we used two versions of *FOM* definition. The version 1 $FOM_{V1}$ consists of a transmission term $FOM_T$ and a phase term $FOM_{phase}$, which are respectively given by

$$FOM_T = \frac{-10\log_{10}\left(1-\sqrt{\sum_{i=3}^{6}\left(|S_{i1}|^2-25\%\right)}\right)-10\log_{10}\left(1-\sqrt{\sum_{i=3}^{6}\left(|S_{i2}|^2-25\%\right)}\right)}{2}, \quad (S1)$$

$$FOM_{Phase} = \min\left\{-10\log_{10}(1-\|[\theta_3-\theta_6, \theta_4-\theta_6, \theta_5-\theta_6]-V_p\|)\right\}, \quad (S2)$$

where $V_p$ is the reference phase vector, $V_p \in \{[-90°, 90°, 180°], [-90°, 180°, 90°], [90°, -90°, 180°], [90°, 180°, -90°], [180°, -90°, 90°], [180°, 90°, -90°]\}$. One has $FOM_{V1}=FOM_T+FOM_{phase}$. Here only the central wavelength of 1.55 μm is considered. In an improved version, the excess loss term $FOM_{EL}$ and the imbalance term $FOM_{IM}$ are used to replace $FOM_T$:

$$FOM_{EL} = -10\log_{10}\frac{\sum_{i=3}^{6}\left(|S_{i1}|^2+|S_{i2}|^2\right)}{2}, \quad (S3)$$

$$FOM_{IM} = -10\log_{10}\left(\frac{\min\{|S_{ij}|^2\}}{\max\{|S_{ij}|^2\}}\right), \quad (i=3,4,5,6, j=1,2). \quad (S4)$$

The total *FOM* is $FOM_{total}=FOM_{EL}+FOM_{IM}+FOM_{phase}$ for a single wavelength. The final *FOM* is given by the mean of $FOM_{total}$ at five wavelength points: $FOM_{V2} = \overline{FOM_{total}(\lambda)}$, $\lambda_i \in \{1530, 1540, 1550, 1560, 1570\}$nm. The design flow starts from a regular linear initial structure by using an EME method. The EME method can obtain the transfer matrix by one-run costing only 36 seconds for single wavelength. To get more accurate result, the FDTD solver was used latter. In the fast-simulation-mesh, there are 10 mesh points per effective-wavelength scale. The simulation of one sample consisting of two runs of the FDTD solver costs ~1.63 minutes. In the high-accuracy-mesh, there are 14 mesh points per effective-wavelength scale, and the local mesh refinements were adopted. In this case, the total simulation time of one sample is ~13.23 minutes. The performance confirmation of the final designed device was carried by FDTD solver with very dense meshes (34 mesh points per effective-wavelength scale); The simulation results are shown in Fig. 4(b)-4(d).

First, the air-slot device was designed. After total 299 generations of iterations, the *FOM* reaches 0.392 dB within total simulation time of ~199.4 hours. Then a SiO$_2$-filled device was designed using this result as initial sample. The design flow includes total 165 generations of iterations, costing total simulation time of ~278.83 hours. The final *FOM* is 1.189 dB.

The geometric parameters of the final device are given in Table S3. As Fig. 2(a) shows, $d_{mn}$ denotes the spacings in *yz*-cross-sections of the optimization region. The final device has a 64-section optimization region with 65 cross-sections. $l_1 \sim l_{63}$, the length of the optimization region $l_{64}$ is fixed to 10 μm. Since there are two input waveguides, the parameter $d_{in}$ is added to denote the axle wire spacing of the input waveguides (Fig. 4(a)). The parameters {$w_{in}$, $d_{in0}$, $d_{in}$, $w_{out}$, $d_{out0}$, $d_{out}$, $l_{in}$, $l_{out}$} are included in the optimization data-set. The SWG period and fill factor are respectively *P*=0.2 μm and *ff*=0.5.

**Table S3 The detail geometric parameters of the final designed SiO$_2$-filled 90° hybrid (unit: μm).**

| $d_{mn}$ | | Optimization Region Parameters in y-direction | | | | | | | | Optimization Region Parameters in x-direction | | Else parameters | |
|---|---|---|---|---|---|---|---|---|---|---|---|---|---|
| | | *n* | | | | | | | | | | | |
| | | 1 | 2 | 3 | 4 | 5 | 6 | 7 | 8 | | | | |
| *m* | 1 | 1.791 | 0.631 | 0.141 | 0.952 | 0.232 | 0.764 | 0.192 | 0.794 | | | | |
| | 2 | 1.773 | 0.643 | 0.160 | 0.907 | 0.211 | 0.756 | 0.204 | 0.774 | $l_1$ | 0.154 | $w_{in}$ | 0.986 |
| | 3 | 1.758 | 0.641 | 0.182 | 0.866 | 0.197 | 0.739 | 0.209 | 0.769 | $l_2$ | 0.310 | $d_{in0}$ | 2.494 |
| | 4 | 1.739 | 0.653 | 0.199 | 0.832 | 0.180 | 0.728 | 0.219 | 0.758 | $l_3$ | 0.471 | $d_{in}$ | 1.999 |
| | 5 | 1.724 | 0.664 | 0.215 | 0.792 | 0.160 | 0.716 | 0.224 | 0.739 | $l_4$ | 0.633 | $w_{out}$ | 0.526 |
| | 6 | 1.701 | 0.679 | 0.239 | 0.752 | 0.143 | 0.724 | 0.245 | 0.732 | $l_5$ | 0.790 | $d_{out0}$ | 1.382 |
| | 7 | 1.689 | 0.693 | 0.266 | 0.709 | 0.128 | 0.730 | 0.257 | 0.706 | $l_6$ | 0.949 | $d_{out}$ | 1.396 |
| | 8 | 1.672 | 0.721 | 0.289 | 0.667 | 0.107 | 0.736 | 0.273 | 0.687 | $l_7$ | 1.106 | $l_{in}$ | 2.018 |
| | 9 | 1.652 | 0.743 | 0.310 | 0.625 | 0.090 | 0.732 | 0.287 | 0.661 | $l_8$ | 1.260 | $l_{out}$ | 1.012 |
| | 10 | 1.641 | 0.743 | 0.321 | 0.604 | 0.090 | 0.724 | 0.292 | 0.668 | $l_9$ | 1.426 | | |
| | 11 | 1.630 | 0.739 | 0.334 | 0.582 | 0.088 | 0.712 | 0.301 | 0.666 | $l_{10}$ | 1.588 | | |
| | 12 | 1.623 | 0.742 | 0.348 | 0.559 | 0.083 | 0.707 | 0.299 | 0.662 | $l_{11}$ | 1.756 | | |
| | 13 | 1.624 | 0.744 | 0.360 | 0.526 | 0.083 | 0.694 | 0.300 | 0.663 | $l_{12}$ | 1.916 | | |
| | 14 | 1.617 | 0.741 | 0.380 | 0.503 | 0.082 | 0.691 | 0.317 | 0.659 | $l_{13}$ | 2.059 | | |
| | 15 | 1.606 | 0.744 | 0.394 | 0.477 | 0.080 | 0.688 | 0.339 | 0.651 | $l_{14}$ | 2.208 | | |
| | 16 | 1.607 | 0.753 | 0.411 | 0.439 | 0.079 | 0.676 | 0.354 | 0.647 | $l_{15}$ | 2.349 | | |
| | 17 | 1.599 | 0.755 | 0.426 | 0.412 | 0.081 | 0.665 | 0.366 | 0.639 | $l_{16}$ | 2.489 | | |
| | 18 | 1.591 | 0.728 | 0.421 | 0.441 | 0.114 | 0.670 | 0.399 | 0.640 | $l_{17}$ | 2.650 | | |
| | 19 | 1.593 | 0.710 | 0.415 | 0.469 | 0.154 | 0.679 | 0.431 | 0.642 | $l_{18}$ | 2.811 | | |
| | 20 | 1.592 | 0.689 | 0.410 | 0.504 | 0.181 | 0.684 | 0.452 | 0.646 | $l_{19}$ | 2.963 | | |
| | 21 | 1.592 | 0.669 | 0.402 | 0.534 | 0.228 | 0.696 | 0.476 | 0.641 | $l_{20}$ | 3.118 | | |
| | 22 | 1.588 | 0.656 | 0.392 | 0.560 | 0.261 | 0.692 | 0.508 | 0.658 | $l_{21}$ | 3.277 | | |

| | | | | | | | | | | | |
|---|---|---|---|---|---|---|---|---|---|---|---|
| 23 | 1.579 | 0.634 | 0.384 | 0.592 | 0.290 | 0.694 | 0.533 | 0.672 | $l_{22}$ | 3.427 | | |
| 24 | 1.581 | 0.615 | 0.383 | 0.627 | 0.320 | 0.688 | 0.561 | 0.673 | $l_{23}$ | 3.573 | | |
| 25 | 1.573 | 0.589 | 0.385 | 0.665 | 0.352 | 0.686 | 0.586 | 0.690 | $l_{24}$ | 3.722 | | |
| 26 | 1.537 | 0.585 | 0.408 | 0.682 | 0.400 | 0.686 | 0.587 | 0.699 | $l_{25}$ | 3.881 | | |
| 27 | 1.502 | 0.577 | 0.431 | 0.703 | 0.443 | 0.689 | 0.592 | 0.710 | $l_{26}$ | 4.021 | | |
| 28 | 1.461 | 0.567 | 0.459 | 0.731 | 0.508 | 0.685 | 0.591 | 0.711 | $l_{27}$ | 4.179 | | |
| 29 | 1.418 | 0.551 | 0.488 | 0.752 | 0.556 | 0.672 | 0.596 | 0.713 | $l_{28}$ | 4.328 | | |
| 30 | 1.384 | 0.534 | 0.509 | 0.790 | 0.605 | 0.660 | 0.596 | 0.721 | $l_{29}$ | 4.492 | | |
| 31 | 1.353 | 0.517 | 0.521 | 0.820 | 0.642 | 0.647 | 0.597 | 0.727 | $l_{30}$ | 4.657 | | |
| 32 | 1.324 | 0.498 | 0.541 | 0.849 | 0.692 | 0.642 | 0.588 | 0.726 | $l_{31}$ | 4.812 | | |
| 33 | 1.287 | 0.476 | 0.556 | 0.880 | 0.736 | 0.627 | 0.588 | 0.727 | $l_{32}$ | 4.971 | | |
| 34 | 1.303 | 0.474 | 0.583 | 0.841 | 0.730 | 0.616 | 0.590 | 0.717 | $l_{33}$ | 5.127 | | |
| 35 | 1.320 | 0.472 | 0.605 | 0.809 | 0.713 | 0.605 | 0.600 | 0.699 | $l_{34}$ | 5.293 | | |
| 36 | 1.335 | 0.481 | 0.632 | 0.777 | 0.697 | 0.599 | 0.611 | 0.688 | $l_{35}$ | 5.446 | | |
| 37 | 1.349 | 0.474 | 0.658 | 0.742 | 0.687 | 0.591 | 0.621 | 0.672 | $l_{36}$ | 5.604 | | |
| 38 | 1.369 | 0.474 | 0.669 | 0.717 | 0.673 | 0.565 | 0.635 | 0.641 | $l_{37}$ | 5.768 | | |
| 39 | 1.394 | 0.464 | 0.679 | 0.676 | 0.665 | 0.543 | 0.656 | 0.610 | $l_{38}$ | 5.934 | | |
| 40 | 1.425 | 0.458 | 0.705 | 0.641 | 0.646 | 0.531 | 0.676 | 0.589 | $l_{39}$ | 6.095 | | |
| 41 | 1.447 | 0.453 | 0.719 | 0.615 | 0.635 | 0.509 | 0.705 | 0.558 | $l_{40}$ | 6.249 | | |
| 42 | 1.443 | 0.458 | 0.762 | 0.588 | 0.621 | 0.523 | 0.704 | 0.562 | $l_{41}$ | 6.394 | | |
| 43 | 1.438 | 0.467 | 0.807 | 0.557 | 0.615 | 0.521 | 0.711 | 0.569 | $l_{42}$ | 6.541 | | |
| 44 | 1.435 | 0.470 | 0.851 | 0.524 | 0.605 | 0.526 | 0.711 | 0.568 | $l_{43}$ | 6.691 | | |
| 45 | 1.424 | 0.479 | 0.890 | 0.498 | 0.604 | 0.518 | 0.720 | 0.575 | $l_{44}$ | 6.843 | | |
| 46 | 1.406 | 0.493 | 0.930 | 0.460 | 0.601 | 0.530 | 0.720 | 0.576 | $l_{45}$ | 7.005 | | |
| 47 | 1.392 | 0.506 | 0.962 | 0.418 | 0.601 | 0.534 | 0.733 | 0.575 | $l_{46}$ | 7.168 | | |
| 48 | 1.366 | 0.530 | 0.998 | 0.381 | 0.594 | 0.550 | 0.745 | 0.583 | $l_{47}$ | 7.337 | | |
| 49 | 1.350 | 0.549 | 1.027 | 0.353 | 0.596 | 0.553 | 0.757 | 0.589 | $l_{48}$ | 7.502 | | |
| 50 | 1.343 | 0.540 | 1.028 | 0.358 | 0.625 | 0.533 | 0.738 | 0.554 | $l_{49}$ | 7.655 | | |
| 51 | 1.343 | 0.537 | 1.020 | 0.379 | 0.655 | 0.509 | 0.723 | 0.523 | $l_{50}$ | 7.811 | | |
| 52 | 1.340 | 0.522 | 1.004 | 0.394 | 0.674 | 0.482 | 0.699 | 0.502 | $l_{51}$ | 7.963 | | |
| 53 | 1.343 | 0.511 | 0.990 | 0.401 | 0.694 | 0.453 | 0.673 | 0.469 | $l_{52}$ | 8.124 | | |
| 54 | 1.342 | 0.512 | 0.975 | 0.411 | 0.719 | 0.440 | 0.666 | 0.443 | $l_{53}$ | 8.282 | | |
| 55 | 1.337 | 0.513 | 0.956 | 0.414 | 0.745 | 0.422 | 0.655 | 0.415 | $l_{54}$ | 8.440 | | |
| 56 | 1.333 | 0.509 | 0.938 | 0.429 | 0.766 | 0.415 | 0.648 | 0.391 | $l_{55}$ | 8.598 | | |
| 57 | 1.331 | 0.513 | 0.924 | 0.438 | 0.787 | 0.401 | 0.633 | 0.367 | $l_{56}$ | 8.753 | | |
| 58 | 1.306 | 0.504 | 0.923 | 0.472 | 0.809 | 0.393 | 0.643 | 0.367 | $l_{57}$ | 8.916 | | |
| 59 | 1.298 | 0.487 | 0.917 | 0.500 | 0.821 | 0.388 | 0.643 | 0.358 | $l_{58}$ | 9.077 | | |
| 60 | 1.277 | 0.477 | 0.912 | 0.528 | 0.842 | 0.371 | 0.645 | 0.348 | $l_{59}$ | 9.233 | | |
| 61 | 1.265 | 0.473 | 0.908 | 0.565 | 0.856 | 0.358 | 0.647 | 0.342 | $l_{60}$ | 9.387 | | |
| 62 | 1.242 | 0.458 | 0.902 | 0.586 | 0.868 | 0.362 | 0.642 | 0.337 | $l_{61}$ | 9.543 | | |
| 63 | 1.225 | 0.445 | 0.898 | 0.601 | 0.890 | 0.369 | 0.646 | 0.330 | $l_{62}$ | 9.693 | | |
| 64 | 1.204 | 0.438 | 0.889 | 0.621 | 0.906 | 0.375 | 0.644 | 0.317 | $l_{63}$ | 9.845 | | |
| 65 | 1.185 | 0.423 | 0.889 | 0.641 | 0.933 | 0.376 | 0.646 | 0.303 | $l_{64}$ | 10.000 | | |

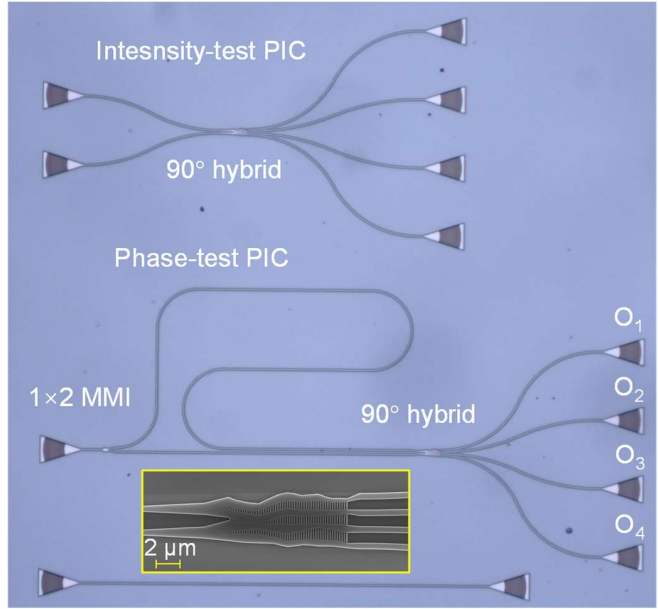

**Fig. S1** The fabricated PICs for SiO$_2$-filled 90° hybrid. Microscope picture for the fabrication silicon intensity-test PIC and phase-test PIC with the SiO$_2$-filled 90° hybrid, inset: SEM image of a 90° hybrid.

### Section S3. The details of the inverse-designed two-channel wavelength demultiplexer

The design details are introduced as follows. The *FOM* is initially given by

$$FOM = \frac{-10\log_{10}\left[\prod_{i=1}^{a}|S_{21}(\lambda_i)|^2 \prod_{j=1}^{b}|S_{31}(\lambda_j)|^2\right]}{a+b}, \quad (S5)$$

$\lambda_i \in \{1470, 1490, 1510, 1530, 1550, 1570\}$ nm, $a=6$, $\lambda_j \in \{1290, 1310, 1330\}$ nm, $b=3$. To obtain the high extinction ratio performance, we introduce a crosstalk term to *FOM*,

$$FOM = \frac{-10\log_{10}\left[\prod_{i=1}^{a}|S_{21}(\lambda_i)|^2 \prod_{j=1}^{b}|S_{31}(\lambda_j)|^2\right]}{a+b} + \frac{-10\log_{10}\left[\prod_{i=1}^{a}(1-|S_{31}(\lambda_i)|^2) \prod_{j=1}^{b}(1-|S_{21}(\lambda_j)|^2)\right]}{a+b}, \quad (S6)$$

Only 3D-FDTD solver is used for electromagnetic (EM) numerical simulation. For each sample, the mode coupling coefficients at different wavelengths can be obtained by one-run of the FDTD solver. Through the design flow, two mesh configurations were used. In the fast-simulation-mesh, there are 10 mesh points per effective-wavelength scale. The simulation of one sample consisting of two runs of EM solver costs ~1 minutes. In the high-accuracy-mesh, there are 18 mesh points per effective-wavelength scale, and the local mesh refinements were adopted. In this case, the simulation of one sample costs ~10.5 minutes. The performance confirmation of the final designed device was carried by FDTD solver with very dense meshes (26 mesh points per effective-wavelength scale with local mesh refinements); The simulation results are shown in Fig. 6(b)-6(c).

The design starts from an initial symmetric Y-branch structure. First, the air-slot device was designed. The optimization region length $L$ is initially set to 6 μm at the beginning of the design flow, and manually is changed to 9 μm by a structure extension manipulation. The final designed $SiO_2$-filled device has $L$ of 10.457 μm. The SWG period $P$ was once set as a variable in optimization data-set, and finally set to fixed value of 0.184 μm. The SWG fill factor is $ff$=0.5.

After 380 generations of iterations, the $FOM$ reaches 0.87 dB within total simulation time of ~ 93.6 hours. Then a $SiO_2$-filled device was designed using this result as initial sample. The design flow includes 76 generations of iterations, costing total simulation time of ~ 209.0 hours. The final $FOM$ is 0.687 dB. The geometric parameters of the final device are given in Table S4.

**Table S4 The detail geometric parameters of the final designed $SiO_2$-filled wavelength demultiplexer (unit: μm).**

| $d_{mn}$ | | Optimization Region Parameters in y-direction | | | | Optimization Region Parameters in x-direction | | Else parameters | |
|---|---|---|---|---|---|---|---|---|---|
| | | n | | | | | | | |
| | | 1 | 2 | 3 | 4 | | | | |
| m | 1 | 1.371 | 0.953 | 0.267 | 0.459 | | | | |
| | 2 | 1.307 | 0.916 | 0.280 | 0.414 | $l_1$ | 0.152 | $w_{in}$ | 1.4 |
| | 3 | 1.246 | 0.887 | 0.286 | 0.377 | $l_2$ | 0.292 | $d_{in}$ | 2.006 |
| | 4 | 1.194 | 0.809 | 0.299 | 0.433 | $l_3$ | 0.536 | $w_{out}$ | 0.5 |
| | 5 | 1.106 | 0.809 | 0.281 | 0.454 | $l_4$ | 0.692 | $d_{out0}$ | 1.303 |
| | 6 | 1.027 | 0.785 | 0.257 | 0.459 | $l_5$ | 0.853 | $d_{out}$ | 1.699 |
| | 7 | 0.973 | 0.740 | 0.226 | 0.494 | $l_6$ | 1.026 | $l_{in}$ | 1.0 |
| | 8 | 0.906 | 0.708 | 0.162 | 0.546 | $l_7$ | 1.195 | $l_{out}$ | 1.0 |
| | 9 | 0.859 | 0.695 | 0.178 | 0.584 | $l_8$ | 1.339 | | |
| | 10 | 0.812 | 0.681 | 0.183 | 0.624 | $l_9$ | 1.490 | | |
| | 11 | 0.781 | 0.684 | 0.233 | 0.680 | $l_{10}$ | 1.636 | | |
| | 12 | 0.744 | 0.693 | 0.265 | 0.726 | $l_{11}$ | 1.788 | | |
| | 13 | 0.710 | 0.684 | 0.336 | 0.694 | $l_{12}$ | 1.923 | | |
| | 14 | 0.676 | 0.679 | 0.404 | 0.681 | $l_{13}$ | 2.061 | | |
| | 15 | 0.640 | 0.681 | 0.482 | 0.653 | $l_{14}$ | 2.183 | | |
| | 16 | 0.611 | 0.678 | 0.549 | 0.627 | $l_{15}$ | 2.310 | | |
| | 17 | 0.559 | 0.678 | 0.571 | 0.669 | $l_{16}$ | 2.466 | | |

| 18 | 0.511 | 0.680 | 0.594 | 0.707 | $l_{17}$ | 2.613 | | |
|---|---|---|---|---|---|---|---|---|
| 19 | 0.463 | 0.686 | 0.629 | 0.763 | $l_{18}$ | 2.773 | | |
| 20 | 0.419 | 0.683 | 0.660 | 0.819 | $l_{19}$ | 2.926 | | |
| 21 | 0.399 | 0.686 | 0.720 | 0.793 | $l_{20}$ | 3.077 | | |
| 22 | 0.384 | 0.671 | 0.776 | 0.774 | $l_{21}$ | 3.232 | | |
| 23 | 0.380 | 0.654 | 0.842 | 0.751 | $l_{22}$ | 3.392 | | |
| 24 | 0.374 | 0.646 | 0.906 | 0.741 | $l_{23}$ | 3.549 | | |
| 25 | 0.434 | 0.621 | 0.947 | 0.706 | $l_{24}$ | 3.741 | | |
| 26 | 0.506 | 0.606 | 0.975 | 0.679 | $l_{25}$ | 3.925 | | |
| 27 | 0.557 | 0.557 | 1.033 | 0.654 | $l_{26}$ | 4.100 | | |
| 28 | 0.624 | 0.521 | 1.067 | 0.627 | $l_{27}$ | 4.277 | | |
| 29 | 0.653 | 0.505 | 1.110 | 0.599 | $l_{28}$ | 4.487 | | |
| 30 | 0.692 | 0.480 | 1.157 | 0.575 | $l_{29}$ | 4.686 | | |
| 31 | 0.741 | 0.459 | 1.171 | 0.558 | $l_{30}$ | 4.875 | | |
| 32 | 0.786 | 0.437 | 1.203 | 0.537 | $l_{31}$ | 5.068 | | |
| 33 | 0.887 | 0.389 | 1.187 | 0.541 | $l_{32}$ | 5.257 | | |
| 34 | 0.986 | 0.328 | 1.171 | 0.553 | $l_{33}$ | 5.452 | | |
| 35 | 1.078 | 0.282 | 1.134 | 0.549 | $l_{34}$ | 5.639 | | |
| 36 | 1.165 | 0.225 | 1.107 | 0.564 | $l_{35}$ | 5.824 | | |
| 37 | 1.206 | 0.231 | 1.114 | 0.558 | $l_{36}$ | 6.013 | | |
| 38 | 1.243 | 0.234 | 1.121 | 0.563 | $l_{37}$ | 6.198 | | |
| 39 | 1.282 | 0.257 | 1.116 | 0.568 | $l_{38}$ | 6.381 | | |
| 40 | 1.309 | 0.259 | 1.113 | 0.583 | $l_{39}$ | 6.568 | | |
| 41 | 1.310 | 0.262 | 1.125 | 0.587 | $l_{40}$ | 6.708 | | |
| 42 | 1.310 | 0.265 | 1.143 | 0.598 | $l_{41}$ | 6.846 | | |
| 43 | 1.287 | 0.263 | 1.160 | 0.596 | $l_{42}$ | 6.987 | | |
| 44 | 1.275 | 0.271 | 1.182 | 0.607 | $l_{43}$ | 7.133 | | |
| 45 | 1.264 | 0.283 | 1.171 | 0.596 | $l_{44}$ | 7.365 | | |
| 46 | 1.273 | 0.302 | 1.129 | 0.589 | $l_{45}$ | 7.605 | | |
| 47 | 1.273 | 0.309 | 1.118 | 0.597 | $l_{46}$ | 7.727 | | |
| 48 | 1.268 | 0.306 | 1.106 | 0.597 | $l_{47}$ | 7.851 | | |
| 49 | 1.278 | 0.299 | 1.086 | 0.625 | $l_{48}$ | 8.105 | | |
| 50 | 1.270 | 0.300 | 1.095 | 0.608 | $l_{49}$ | 8.235 | | |
| 51 | 1.278 | 0.313 | 1.092 | 0.588 | $l_{50}$ | 8.355 | | |
| 52 | 1.297 | 0.311 | 1.101 | 0.573 | $l_{51}$ | 8.486 | | |
| 53 | 1.320 | 0.319 | 1.106 | 0.554 | $l_{52}$ | 8.623 | | |
| 54 | 1.312 | 0.317 | 1.140 | 0.581 | $l_{53}$ | 8.857 | | |
| 55 | 1.302 | 0.309 | 1.158 | 0.620 | $l_{54}$ | 9.090 | | |
| 56 | 1.292 | 0.303 | 1.188 | 0.637 | $l_{55}$ | 9.318 | | |
| 57 | 1.276 | 0.300 | 1.208 | 0.659 | $l_{56}$ | 9.544 | | |
| 58 | 1.263 | 0.310 | 1.198 | 0.653 | $l_{57}$ | 9.767 | | |
| 59 | 1.247 | 0.319 | 1.192 | 0.637 | $l_{58}$ | 9.987 | | |
| 60 | 1.242 | 0.326 | 1.182 | 0.618 | $l_{59}$ | 10.223 | | |
| 61 | 1.224 | 0.329 | 1.157 | 0.602 | $l_{60}$ | 10.457 | | |